\documentclass[10pt, paper=a4, UKenglish]{article}
\usepackage{graphicx}
\usepackage{amsmath}
\usepackage{amsfonts}
\usepackage{multirow}
\usepackage{stmaryrd}
\usepackage{placeins}
\usepackage{cite}

\def\Title#1{\begin{center} {\Large #1 } \end{center}}
\def\Author#1{\begin{center}{ \sc #1} \end{center}}
\def\Contribution#1{\begin{center}{#1} \end{center}}
\def\Address#1{\begin{center}{ \it #1} \end{center}}

\newenvironment{Abstract}{\begin{quotation} \begin{center}
             \large ABSTRACT \end{center}\bigskip 
      \begin{large}}{\end{large} \end{quotation}}

\def\Acknowledgements{\bigskip  \bigskip \begin{center} \begin{large}
      \bf ACKNOWLEDGEMENTS \end{large}\end{center}}

\usepackage{color}
\usepackage{lineno}
\usepackage{subfig}
\usepackage{hyperref}

\begin{document}

\large
\begin{titlepage}

\vfill
\Title{Graph Neural Network-Based Track Finding in the LHCb Vertex Detector}
\vfill

\Author{Anthony~Correia$^1$, Fotis~I.~Giasemis$^{2,1}$\textasteriskcentered, Nabil~Garroum$^1$, Vladimir~Vava~Gligorov$^{1,3}$, Bertrand Granado$^2$}

\Address{$^1$ LPNHE, Sorbonne Université, CNRS/IN2P3, Paris, France}
\Address{$^2$ LIP6, Sorbonne Université, CNRS, Paris, France}
\Address{$^3$ Organisation Européenne pour la Recherche Nucléaire (CERN)}
\Contribution{\textasteriskcentered \, Corresponding author: \href{mailto:Fotis.Giasemis@cern.ch}{Fotis.Giasemis@cern.ch}}
\vfill

\begin{Abstract}
The next decade will see an order of magnitude increase in data collected by high-energy physics experiments, 
driven by the High-Luminosity LHC (HL-LHC). The reconstruction of charged particle trajectories (tracks) has
always been a critical part of offline data processing pipelines. The complexity of HL-LHC data will however 
increasingly mandate track finding in all stages of an experiment's real-time processing. 
This paper presents a GNN-based track-finding pipeline tailored for the Run 3 LHCb 
experiment's vertex detector and benchmarks its physics performance and computational cost against existing classical
algorithms on GPU architectures. A novelty of our work compared to existing GNN tracking pipelines is batched execution,
in which the GPU evaluates the pipeline on hundreds of events in parallel. We evaluate the impact of neural-network 
quantisation on physics and computational performance, and comment on the outlook for GNN tracking algorithms for
other parts of the LHCb track-finding pipeline.
\end{Abstract}

\vfill

\end{titlepage}

\normalsize 


\section{Introduction}
\label{intro}

Modern high-energy physics (HEP) collider experiments rely on a precise reconstruction of charged particle trajectories (tracks) 
to identify physics processes of interest and to separate them from backgrounds. As the instantaneous luminosity at which 
experiments operate increases so does the number and density of hits in each individual ``event'' and therefore the track-finding
complexity. At the same time this increasing complexity makes it more essential to perform tracking at the earliest stages of
the processing pipeline, including in real-time. The LHCb~\cite{Aaij:2019zbu,LHCb:2020kay} and ALICE~\cite{ALICE:2011nnz, Rohr:2017ydv, Eulisse:2024qfu} detectors at the Large Hadron Collider (LHC)
already perform tracking\footnote{This is a partial track reconstruction in LHCb's case, and a full one in the case of ALICE.} 
in software at the full LHC collision rate. The ATLAS~\cite{CERN-LHCC-2017-020} and CMS~\cite{Collaboration:2759072} collaborations are currently building upgraded 
detectors which will be able to operate at the High-Luminosity LHC (HL-LHC), at instantaneous luminosities up to four times 
higher than the current ATLAS and CMS detectors and almost forty times higher than the current LHCb detector.
At the HL-LHC both ATLAS and CMS will increase the rate at which track finding is performed in software by roughly an order of magnitude compared to the present, while CMS will additionally perform~\cite{CERN-LHCC-2020-004} a partial track finding using FPGA cards at the full LHC collision rate. 

While tracking is a major part of the overall computational budget for all these four experiments, the computational 
cost of classical tracking algorithms scales with the number of hits to the power 2-3~\cite{CamporaPerez:2021jhc, Fruhwirth:1990zas} for
the same physics performance. At the same time, the development of computing architectures is increasingly driven by machine-learning and artificial intelligence applications, such as the deployment of specialized hardware like Google's tensor processing units and Nvidia's tensor cores inside GPUs for efficient deep learning computations. While all major high-energy physics experiments have successfully  reoptimized their classical tracking algorithms to 
efficiently exploit a range of current parallel computer architectures over the past decade, it is also worth asking whether tracking algorithms
based on neural networks might be a better long-term fit to the hardware which our reconstruction has to run on. 

There has been a significant amount of R\&D~\cite{Golling:2018yyi,Calafiura:2018cdd,Amrouche:2019wmx,Amrouche:2021nbs,ExaTrkX:2020apx,Caillou:2024dcn} performed in the community on this topic over the last years. In particular, 
the Exa.TrkX collaboration developed a highly parallelizable Graph Neural Network (GNN)-based pipeline for track 
finding~\cite{ExaTrkX:2021abe}. This pipeline was initially designed for $4\pi$ tracking detectors in a magnetic field, 
similar to those used in ATLAS and CMS, specifically for the High-Luminosity upgrade of the LHC (HL-LHC). Using this pipeline
as a starting point, we developed~\cite{etx4veloctd} our own pipeline for track finding in LHCb's vertex detector, called ETX4VELO. 
In this paper, we extend our earlier results~\cite{etx4veloctd} by partially implementing ETX4VELO in C++ and CUDA, excluding the triplet steps described in Section~\ref{pipeline:triplet_building}. Additionally, we present several optimizations, including minimizing the sizes of the ML models, improving the models' architectures, and restricting k-NN computations to successive planes. We report physics and computational performance with and without neural network quantisation and discuss the outlook for further speedups of the ETX4VELO pipeline and applications to track finding in the rest of the LHCb detector.

\section{The LHCb detector and computing framework}

\begin{figure}[!htb]
  \centering
  \includegraphics[width=0.90\linewidth]{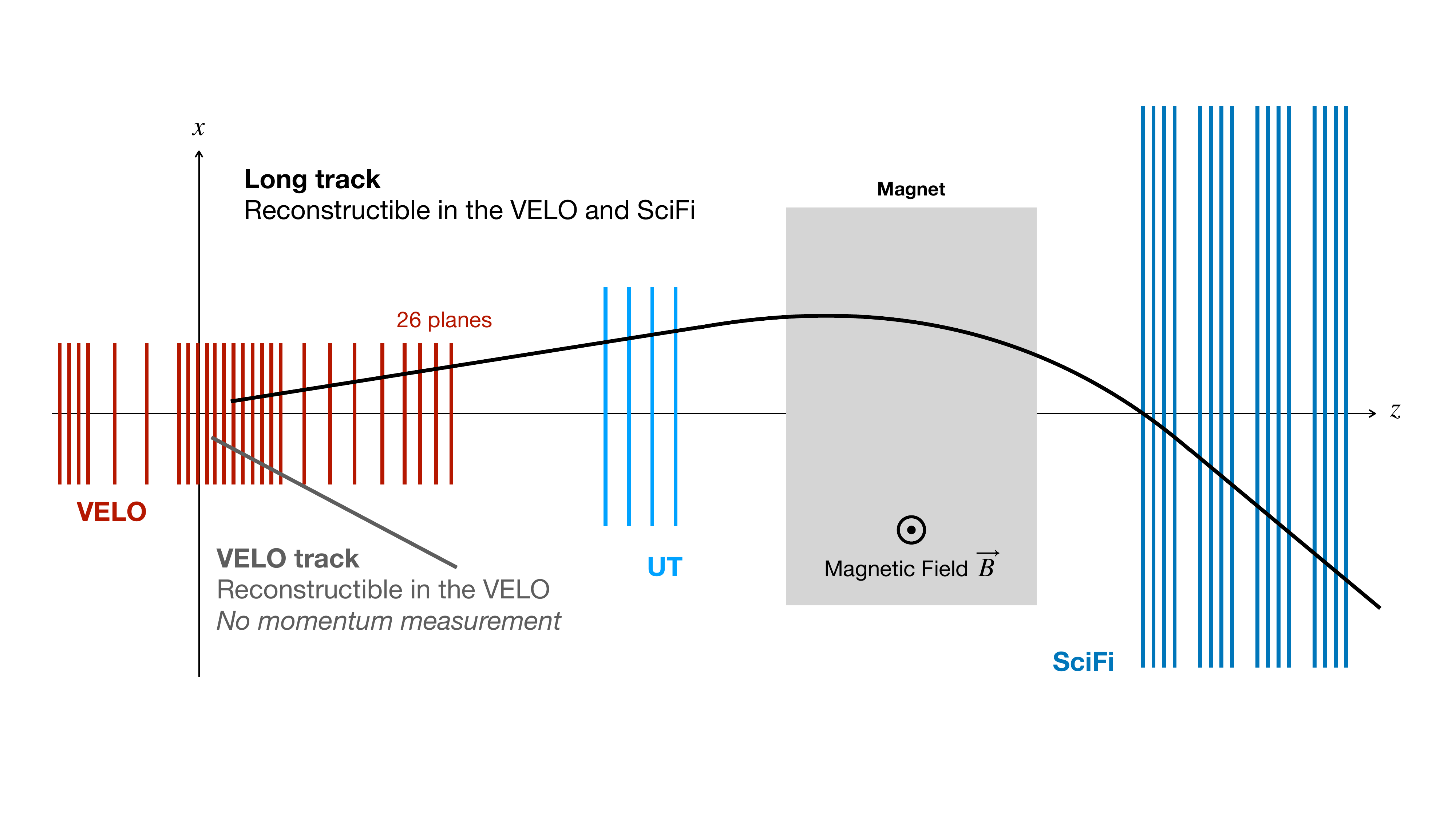}
  \caption{Sketch of the LHCb tracking system in Run~3.}
  \label{fig:tracking_system}
\end{figure}

The LHCb Run~3 detector~\cite{LHCb:2012doh} is a general-purpose detector covering a pseudorapidity ($\eta$) region between 2 and 5 and 
optimised for making precision measurements of heavy flavour hadrons. The LHC proton bunches cross every 25 ns within its 
Vertex Locator (VELO), with a beam-beam collision rate of around 24~MHz in 2024 datataking and a nominal pileup of around five 
proton-proton collisions per LHC bunch crossing. The particles produced in these collisions are detected by the tracking system, 
illustrated in Figure~\ref{fig:tracking_system}. This system consists of:
\begin{itemize}
    \itemsep0em
    \item the \textbf{Vertex Locator (VELO)}~\cite{Bediaga:2013tje}, surrounding the proton-proton interaction region, encompassing $n_{\text{planes}} = 26$ planes of $55\times55$ ${\mu m}^2$ silicon pixels,
    \item the \textbf{Upstream Tracker (UT)}~\cite{LHCb:2014uqj}, placed before the magnet station, including 4 planes of silicon strips, and
    \item the \textbf{Scintillating Fibre Tracker (SciFi)}~\cite{LHCb:2014uqj}, located after the magnet station, composed of 12 planes of scintillating fibres.
\end{itemize}

The LHCb data acquisition can sustain a peak data rate of 5~TB/s which must be reduced to 10~GB/s before being
written to long-term storage. This reduction is performed by a two-stage real-time processing system, the so-called trigger. The first stage is implemented on GPUs and is called Allen\footnote{\url{https://gitlab.cern.ch/lhcb/Allen}}~\cite{LHCb:2020kay}. Allen performs a partial event
reconstruction, including tracking, and reduces the rate to between 70 and 200~GB/s using around 50 physics selections. 
The second stage is implemented on CPUs using the Moore\footnote{\url{https://gitlab.cern.ch/lhcb/Moore}} application based on the
Gaudi\footnote{\url{https://gitlab.cern.ch/gaudi}} framework. Moore performs a full offline-quality reconstruction of the LHCb 
detector and reduces the rate to 10~GB/s using over two thousand physics selections. 

Given this context, the Allen framework and LHCb offer an interesting use-case for developing Neural Network-based algorithms on GPUs, with a specific emphasis on high-throughput performance. In particular LHCb processes the full LHC collision rate using only around 320~GPU cards,
leading to a minimum required throughput of around 80~kHz for the full processing pipeline on a single GPU. This in turn makes it essential to minimize
overheads caused by data transfers between the host server and the GPU, which Allen achieves by processing batches of O(1000) events at
a time. To the best of our knowledge this is the first time that a GNN tracking pipeline has been implemented using batched processing
of this kind, or targeting these kinds of throughputs per GPU.

\section{Track Topologies in the VELO}
\label{track_topology}

As there is no magnetic field in the VELO, the tracks are straight lines. Since the LHCb detector is not symmetric around the 
interaction region, tracks are classified as ``forward'', i.e. travelling towards the LHCb magnet and the rest of the LHCb 
detector, or ``backward'', travelling away from the LHCb detector. When evaluating tracking efficiencies we only consider 
particles which leave at least three hits in the VELO as ``reconstructible''.~\cite{Li:2021oga}

\begin{figure}[!htb]
  \centering
  \includegraphics[width=0.25\linewidth]{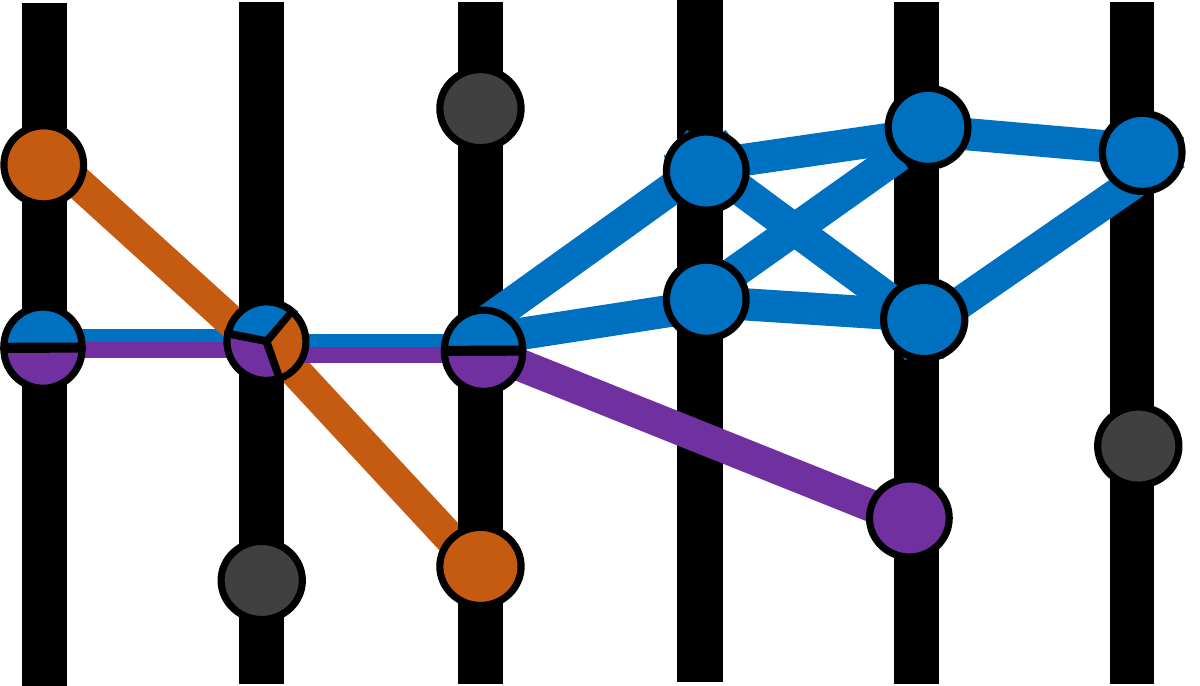}
  \caption{Simplified example of tracks to be reconstructed in the VELO. The blue and purple tracks share three hits prior to diverging. The purple track jumps from plane 3 to 5, missing 4. The orange track intersects the blue and purple tracks. Dark points represent hits unassociated with any particle. When considering the hits as graph nodes, lines between hit nodes represent the genuine edges, as defined in this work. Reproduced from~\cite{etx4veloctd}.}
  \label{fig:tracks}
\end{figure}

A VELO plane consists of four overlapping sensor layers, displaced along the $z$-axis, collaboratively covering the desired acceptance in the $\left(x, y\right)$-plane. The effective hit efficiency of the VELO is around 99\% in simulation, so tracks must be allowed to skip a VELO plane during the
reconstruction. A track can also leave more than one hit per plane, in most cases when it traverses overlapping sensor layers within a
given plane. Material interactions frequently produce positron-electron pairs, resulting in two tracks that initially share hits before 
diverging. In fact 55\% of electrons share hits with another particle, which motivates the need to take particular care when reconstructing them. Two tracks may also accidentally intersect, leading to a shared hit. The tracks that an effective VELO tracking algorithm 
should be capable of reconstructing are illustrated in Figure~\ref{fig:tracks}.

\section{Datasets}
\label{datasets}

In order to have the information required for the training process of the machine learning models involved, such as the truth labels required for calculating the loss for a classification task, the pipeline is trained on simulated events. Similarly, for the evaluation of the accuracy of the algorithm, a simulation sample is necessary in order to have information about the particles produced in each event. 

The results presented in this paper have been obtained using events produced with the full LHCb detector simulation in which $p$-$p$ collisions are generated using \texttt{PYTHIA}~\cite{Sjostrand:2007gs} and decayed using EvtGen~\cite{Lange:2001uf}. The interaction of the generated particles with the detector, and its response, are implemented using the \texttt{Geant4} toolkit~\cite{Allison:2006ve} as described in~\cite{Clemencic:2011zza}. We use simulated minimum bias samples corresponding to typical LHCb datataking conditions between 2022 and 2025, with an average of 5.3 inelastic $p$-$p$ collisions which produce at least one particle with momentum above 2~GeV in the LHCb detector acceptance per event. In our sample, there are, on average, 150 particles in the VELO acceptance, and 2,200 hits, in each event. In the simulation used for this paper 
around 15\% of the hits are spilled over from prior events and therefore noise. The embedding network and GNN, introduced in Section~\ref{pipeline}, are trained on 700,000 events adhering to the selection criteria below.
\begin{enumerate}
    \itemsep0em
    \item \textbf{Removal of non-linear particle tracks}: The hits of the particle tracks that are not sufficiently linear due to multiple scattering (largely soft electrons) are removed. This is assessed by fitting a line to the particle hits and applying an upper limit to the average squared distance between the hits and the line. This criterion improves the physics performance during the training but also excludes 2.5\% of the tracks reconstructible in the VELO.
    \item \textbf{Minimum number of VELO hits}: A minimum of 500 genuine VELO hits is required.
    \item \textbf{Exclusion of tracks with insufficient hits}: Tracks with fewer than 3 hits are excluded.
\end{enumerate}
These criteria are not imposed on the test samples. Allen's existing tracking algorithms, which are 
used by LHCb in 2024 datataking, are used as a reference for both the physics and computational performance of ETX4VELO, described in detail in Sections~\ref{sec:performance} and \ref{sec:gpu-implemenation} respectively.

\section{The ETX4VELO Pipeline}
\label{pipeline}

In general, a graph is a pair $G = (V, E)$, where $V$ is a finite set of vertices (or nodes), and $E$ is the set of connections (known as edges) between these nodes. In our problem, the hits left by charged particles in the VELO detector form a point cloud in 3-dimensional space. However, they can also be represented as a graph, in which the successive hits (or nodes) of each particle traversing the detector are connected to each other. This graph can be thought of as the ``truth graph'' that perfectly describes the VELO event. Conceptually, the end-goal of the ETX4VELO pipeline is to produce a graph that closely approximates this truth graph.

More specifically, the basic idea of our GNN pipeline is to build an initial graph of possible connections between hits in the detector, accurately classify these connections as correct (true/genuine) or incorrect (fake), and then, after discarding the fake ones, transform them into a set of track objects, containers of the hits of each track, which can then be understood and used by the rest of the algorithms in LHCb's real-time pipeline. Since a graph with N nodes consisting of all possible node connections would have $C(N,2) = N(N-1)/2$ edges, and thus would be prohibitively large, a key challenge is to construct this initial graph in such a way that nearly all initial connections made are part of the final graph, while as many fake connections as possible are not. To illustrate this, for a typical VELO event of $N=2,200$ hits, a maximally connected graph would have $C(N,2)=5 \times 10^6$ edges, while our method, described in the following paragraphs, creates on average 30,000 edges. 

Our pipeline consists of 7 steps. We summarize the steps here, and then give further details on each step in dedicated subsections. Starting with the hits alone, the first two steps of the ETX4VELO pipeline construct a rough graph, $G^{\text{hit}}_{\text{rough}}$. Hits are first embedded into a Euclidean space using a Multi-Layer Perceptron (MLP). The MLP is trained to position hits that are likely to be connected by an edge close to each other in the embedding space. Subsequently, $k$-Nearest Neighbours ($k$-NN) algorithms are applied in the embedding space to collect edges that are most likely to be genuine. Nodes are considered at most two planes apart when identifying edge candidates. 

The next four steps involve the GNN. A novelty of the ETX4VELO pipeline is the classification of edge connections in addition to individual edges, which allows for the separation of tracks that share hits. This involves transitioning from a graph of connected hits, $G^{\text{hit}}$, to a graph of edges, $G^{\text{edge}}$. Edge connections are referred to as triplets because they are formed by three hits. To minimise the number of edge connections, which increases exponentially with the number of edges, it is essential to first filter out fake edges.

First, a GNN encodes each edge in $G^{\text{hit}}_{\text{rough}}$ ($i \to j$) into a high-dimensional vector $e_{i\to j}$, with dual training 
targets: edge classification and triplet classification. Second, the \textit{edge classifier} network transforms these encodings into edge scores ranging from 0 (fake) 
to 1 (genuine). Edges with scores below $s_{\text{edge, min}}$ are discarded, resulting in the purified hit graph
$G^{\text{hit}}_{\text{purified}}$. Third, the graph of edges $G^{\text{edge}}$ is built from the purified hit graph
$G^{\text{hit}}_{\text{purified}}$. Finally, the \textit{triplet classifier} network transforms the pairs of edge encodings 
(triplets) into triplet scores. Triplets scoring below $s_{\text{edge, min}}$ are discarded, resulting in the purified edge graph $G^{\text{edge}}_{\text{purified}}$.

The final step involves constructing tracks from the purified edge graph $G^{\text{edge}}_{\text{purified}}$. The Exa.TrkX pipeline 
applies a Weakly Connected Component (WCC) algorithm~\cite{tarjan} to the purified hit graph $G^{\text{hit}}_{\text{purified}}$ to interpreted 
sets of connected hits as tracks. The classification of edge connections rather than only edges allows the WCC algorithm to be modified 
in such a way as to allow tracks to share multiple hits, which is particularly important for the efficiency of reconstructing
electron-positron pairs.

\begin{figure}[!htb]
  \centering
  \includegraphics[width=0.98\linewidth]{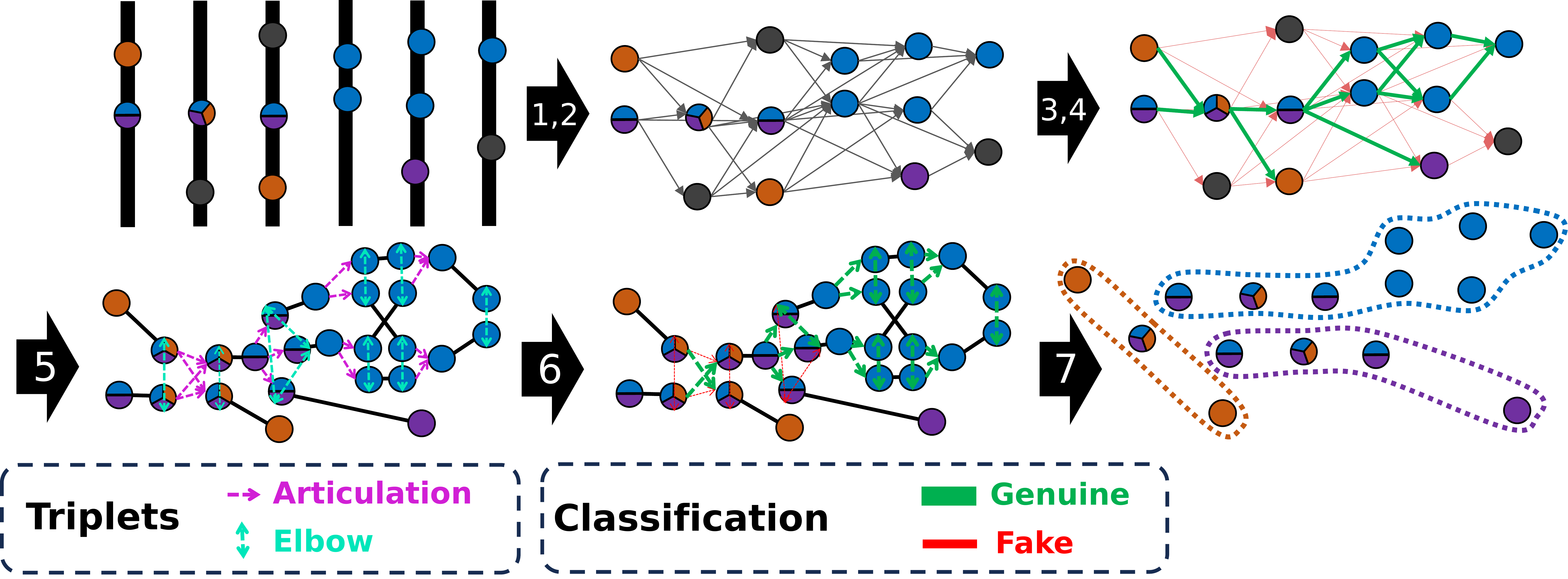}
  \caption{Illustration of the pipeline's 7 steps, beginning with hits from the minimalist example in Figure~\ref{fig:tracks}. Steps entail (1) embedding the hits, (2) building a rough hit graph, (3) encoding the edges, (4) classifying the edges and discarding fakes (in red), (5) constructing the edge graph with edge-to-edge connections called triplets, (6) classifying and removing fake triplets, and finally, (7) producing the tracks.}
  \label{fig:pipeline}
\end{figure}

All these steps are illustrated in Figure~\ref{fig:pipeline}.

\subsection{Hit Embedding and Rough Graph Construction}
\label{pipeline:graph_building}

To construct the graph $G^{\text{hit}}_{\text{rough}}$, one could connect each hit to all hits on the next two planes, allowing for the possibility of a missing plane due to pixel inefficiencies. However, this approach results in an excessive number of edges, thereby increasing the GNN's inference time and memory usage. To enhance throughput, it is essential to minimise the graph size at this stage. We start by describing the operation of our method and then the training process and loss function in Eqs.~\eqref{eq:loss1} and \eqref{eq:loss2}.

Most VELO tracks are produced directly from the initial proton-proton interactions, which occur in a relatively narrow interaction region with a spread of
around 45~mm in $z$ and around 30~$\mu$m in $x$ and $y$. This fact strongly constrains which edges have to be considered when constructing
our graph. The embedding MLP captures this by accepting the hits $h$  as input and embedding them into an $n_{\text{dim}}$-dimensional space $e_h \in \mathbb{R}^{n_{\text{dim}}}$. This is done by passing the normalised\footnote{The inputs of a neural network are normalised using the mean and standard deviation calculated from a representative set of events.} cylindrical $\left(r, \phi, z\right)$ coordinates to the MLP. In this space, likely connected hits are positioned close together based on a reference squared distance $m = 1$\footnote{Using the squared distance instead of the distance avoids the computational cost of square root calculations.}, while unlikely connections are spaced apart. Here, the squared distance $d^2\left(a,b\right)$ between two hits $a$ and $b$ is defined as the usual Euclidean distance
\begin{equation}
    d^2\left(a,b\right) = \lVert e_a - e_b\rVert^2 = \sum_{i=1}^{n_\text{dim}} \left(e_{a,i} - e_{b,i} \right)^2\,.
\end{equation}
Using this trained embedding MLP, a hit on plane $p$ is connected to hits on the next two planes, $p + 1$ and $p + 2$, if they are within 
a squared distance of $d^2_\text{max}$. To avoid an excessive number of edges, a maximum of $k_{\text{max}}$ edges per node is imposed.
Consequently, the rough graph $G^{\text{hit}}_{\text{rough}}$ is constructed by applying a $k_{\text{max}}$-NN algorithm 
on plane $p \in \left\llbracket 0, n_\text{planes} - 1 \right\rrbracket$ to the next two planes, $p + 1$ and $p + 2$, under a maximum 
squared distance of $d^2_\text{max}$. The k-NN implementation from \texttt{faiss}~\cite{Johnson:2021tbd} is used for this purpose. The 
values of the hyperparameters $k_\text{max}$ and $d^2_\text{max}$ are determined post-training to balance the tradeoff between efficiency and clone rate, metrics described in Section~\ref{sec:performance}. Here these values are chosen to be $k_\text{max} = 50$ and $d^2_\text{max} = 0.9$.

In the training process, each step corresponds to one event, with noise hits removed as they are considered random and unrelated. 
The training set $T$ is composed of hit pairs from a query node $q \in Q$ to another node $a$ on the next two planes, representing 
$q \to a$ edge candidates. To focus on significant particles, a hit must belong to a reconstructible particle within acceptance 
and not be an electron to qualify as a query node. Almost all of the edges from electrons are identified by the network without training specifically on them and thus electrons are excluded. The training set $T = T_{\text{genuine}} \cup T_{\text{fake}}$ includes both connected pairs $T_{\text{genuine}}$ 
and disconnected pairs $T_{\text{fake}}$. It is constructed by merging three sets of pairs:
\begin{itemize}
    \itemsep0em
    \item[] \textbf{Hard-negative mining}: Fake pairs are generated using the same $k_{\text{max}}^{\text{training}}$-NN procedure with $(d_{\text{max}}^{\text{training}})^2$ as during inference, representing fake pairs that would be classified as genuine during inference. The values $k_{\text{max}}^{\text{training}} = 50$ and $(d_{\text{max}}^{\text{training}})^2 = 1.5$ are used.
    \item[] \textbf{Random pairs}: For each query point, $n_{\text{random}} = 1$ pairs are included.
    \item[] \textbf{Genuine edges}: All genuine edges from the query points are added to the training set.
\end{itemize}
To train the embedding MLP to reduce the distance of genuine pairs and increase that of fake pairs, the following loss function is minimised:
\begin{equation} \label{eq:loss1}
    \mathcal{L} = \mathcal{L}_{\text{fake}} + w_{\text{genuine}} \times \mathcal{L}_{\text{genuine}}\,,
\end{equation}
where $\mathcal{L}_{\text{genuine}}$ and $\mathcal{L}_{\text{fake}}$ are the normalised pairwise hinge embedding losses~\cite{cortes} for genuine and fake examples, respectively, defined as:
\begin{equation} \label{eq:loss2}
    \mathcal{L_{\text{genuine}}} = \frac{1}{\left|T_{\text{genuine}}\right|} \sum_{(q, a) \in T_{\text{genuine}}} d^2\left(q,a\right)
     \quad\text{and}\quad 
    \mathcal{L_{\text{fake}}} = \frac{1}{\left|T_{\text{fake}}\right|}  \sum_{(q, b) \in T_{\text{fake}}} \max{\left(0, m - d^2\left(q,b\right)\right)}\,,
\end{equation}
The margin $m$, representing a squared distance threshold, is fixed at $m = 1$. The parameter $w_{\text{genuine}} > 1$ reduces the likelihood of excluding true edges within this margin, thus favouring the inclusion of genuine edges over the exclusion of false ones. The values $m$ and $w$, similary to $k_\text{max}$ and $d^2_\text{max}$, are chosen based on an elementary hyperparameter exploration.

\subsection{Graph Neural Network and Classifiers}
\label{pipeline:gnn}

\begin{figure}[!htb]
  \centering
  \includegraphics[width=\linewidth]{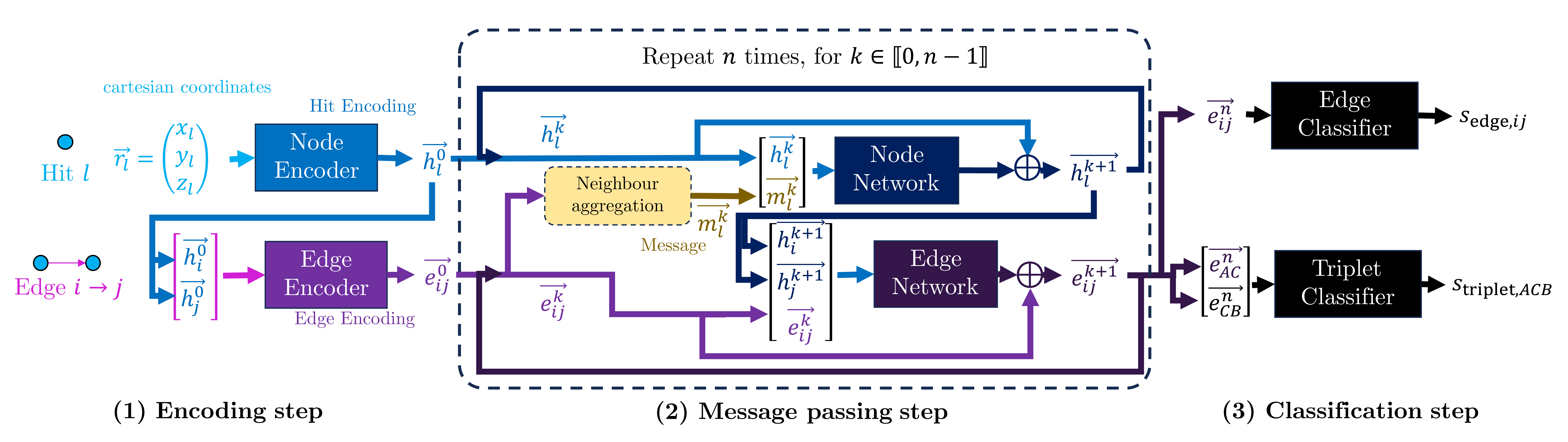}
  \caption{Schematic of the GNN architecture, highlighting: (1) hit and edge encodings, (2) $n$ message passing steps, and (3) subsequent edge and triplet classifications. The node and edge encoders, the node and edge networks and the edge and triplet classifiers are all MLPs.}
  \label{fig:gnn}
\end{figure}

The Graph Neural Network (GNN) is employed to derive edge encodings used for both edge and triplet classification. The architecture of the GNN closely follows that of the Exa.TrkX collaboration, with minor deviations, as illustrated in Figure~\ref{fig:gnn}. The node and edge encoders, the node and edge networks and the edge and triplet classifiers are all MLPs, and should not be confused with the embedding network described in Section~\ref{pipeline:graph_building}, which also happens to be an MLP.

Initially, the hits $l$ are encoded from their normalised Cartesian coordinates $\vec{r}_l = (x_l, y_l, z_l)$ into an $n_h$-dimensional space $\vec{h^0_l} \in \mathbb{R}^{n_h}$ using the node encoder. The concatenated node features $\vec{h^0_i}$ and $\vec{h^0_j}$ of edges $i \to j$ are input to the edge encoder, producing the edge encodings $e^{0}_{ij} \in \mathbb{R}^{n_e}$ in an $n_e$-dimensional space.

The hit and edge encodings are then iteratively updated over $n$ message passing steps. These steps allow the encodings to incorporate information from distant neighbours. During each message passing step $k \in \{0, \ldots, n - 1\}$, a message $m^{k}_l$ is computed for each hit $l$ by aggregating the encodings of the edges connected to and from hit $l$. The message is computed as follows:
\begin{equation}
    m^{k}_l = \left[\sum_{j\text{ s.t. }l \to j\text{ exists}} e^{k}_{lj}, \sum_{i\text{ s.t. }i \to l\text{ exists}} e^{k}_{il}\right]\,
\end{equation}
where $\left[\cdot,\cdot\right]$ denotes concatenation. This operation of aggregating edge encodings by summing them, using terminology from deep learning frameworks, will be referred to as \texttt{scatter\_add}. The node network updates the hit encodings $h^{k+1}_{l}$ using the previous hit encodings $h^{k}_{l}$ and the message $m^{k}_{l}$, incorporating a residual connection. Similarly, the edge encodings are updated to $e^{k+1}_{ij}$ using the previous edge encodings $e^{k}_{ij}$ and the updated hit encodings $h^{k+1}_{i}$ and $h^{k+1}_{j}$, also with a residual connection. 

Edge encodings are sufficient for both the edge classifier and triplet classifier. Therefore, the hit encodings are only 
utilised during the encoding and message passing steps to compute and update the edge encodings. The GNN is trained to classify both 
edges and triplets by minimizing the sum of the edge and triplet losses:\begin{equation}
    \mathcal{L} = \mathcal{L}_{\text{edges}} + \mathcal{L}_{\text{triplets}}\,.
\end{equation}
We use sigmoid focal losses~\cite{Lin:2017fqe} for the edge and triplet classification since, for the purposes of our pipeline, it outperforms the traditional binary cross-entropy loss. To ensure the GNN focuses on relevant triplets, edges with scores below 0.5 are discarded before triplet building and classification.

Several optimizations have been applied to the GNN to improve performance and efficiency. The size of the GNN was significantly reduced, with node and edge encodings now residing in a 32-dimensional space ($n_h = n_e = 32$), down from the initial 256 dimensions. The number of graph iterations was reduced to $n = 5$. Despite the reduction in network size, several changes and corrections were made to maintain reasonable physics performance. Notably, the node and edge networks used at each message passing step are distinct, making the GNN non-recurrent \cite{ctd2023exatrkx}. This approach increases the number of trainable parameters but keeps the throughput unchanged while greatly improving the physics performance. 

\subsection{Triplet Building}
\label{pipeline:triplet_building}

\begin{figure}[!htb]
  \begin{center}
  \subfloat[]{\includegraphics[width=0.16\linewidth]{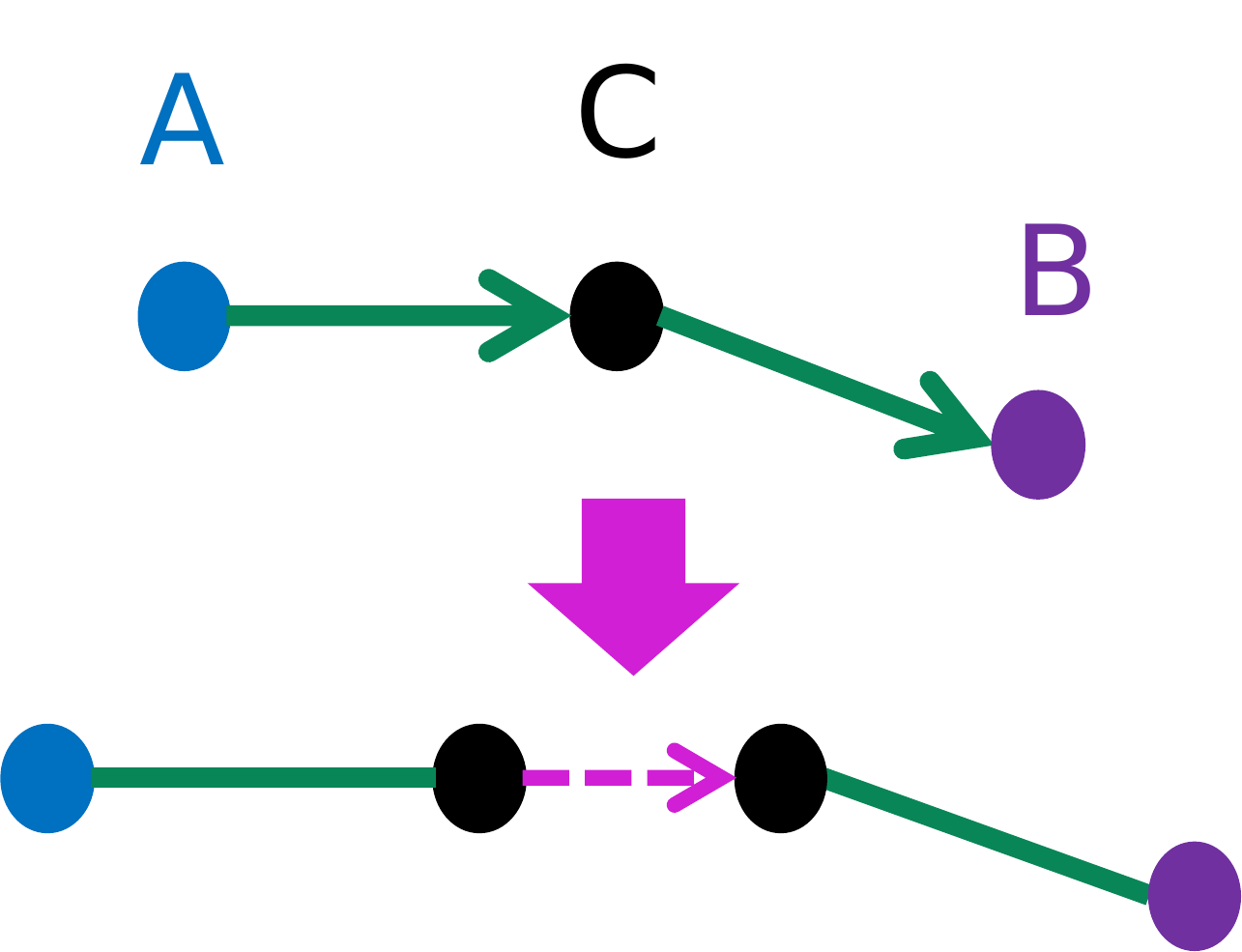}}
  \qquad
  \subfloat[]{\includegraphics[width=0.15\linewidth]{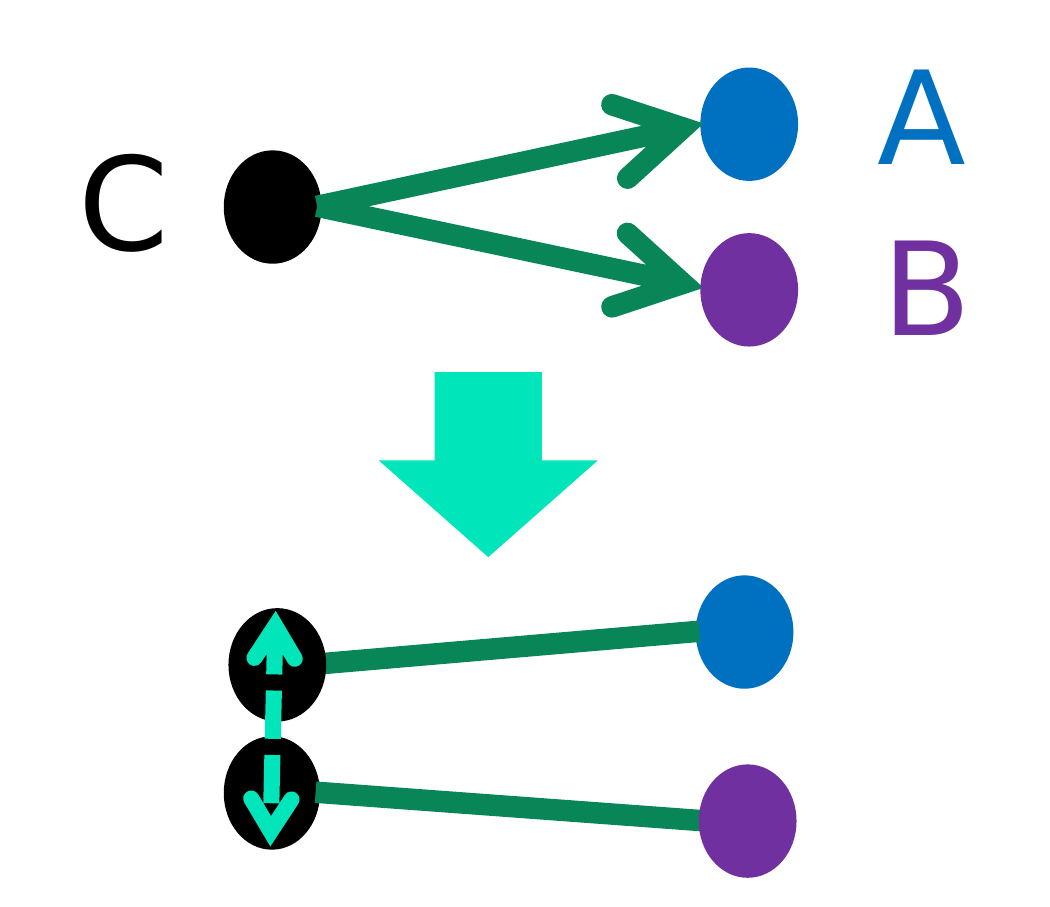}}
  \qquad
  \subfloat[]{\includegraphics[width=0.15\linewidth]{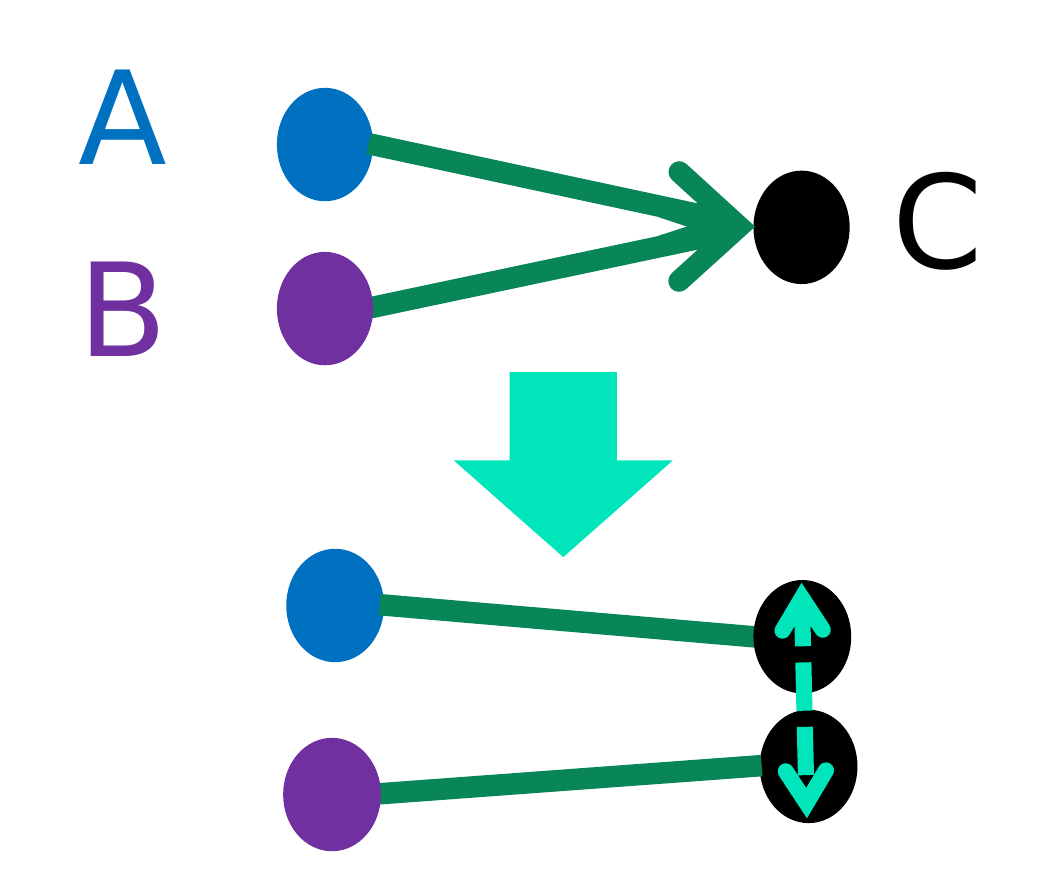}}
  \caption{Visual representation of the three triplet configurations in the edge graph: (a) the articulation, (b) the left elbow and (c) the right elbow.}
  \label{fig:triplets}
  \end{center}
\end{figure}

Edge connections, or triplets~\cite{ExaTrkX:2020apx}, are constructed from the purified hit graph $G^{\text{hit}}_{\text{purified}}$. Each triplet consists of exactly three hits: one common hit $C$ and two other hits, $A$ and $B$. There are only three types of triplets that can be formed, as illustrated in Figure~\ref{fig:triplets}.
\begin{itemize}
    \item[] \textbf{Articulation}: Two consecutive edges, $A \to C$ and $C \to B$, with the common hit in the middle.
    \item[] \textbf{Left Elbow}: Edges $C \to A$ and $C \to B$, with the common hit on the left.
    \item[] \textbf{Right Elbow}: Edges $A \to C$ and $B \to C$, with the common hit on the right.
\end{itemize}
It was found that using separate triplet classifiers for articulations and elbows led to better performance. Each of these classifiers is an MLP ending with a layer with 1 unit, and a sigmoid activation function.

\subsection{Track Building}
\label{pipeline:track_building}

\begin{figure}[!htb]
  \centering
  \includegraphics[width=0.8\linewidth]{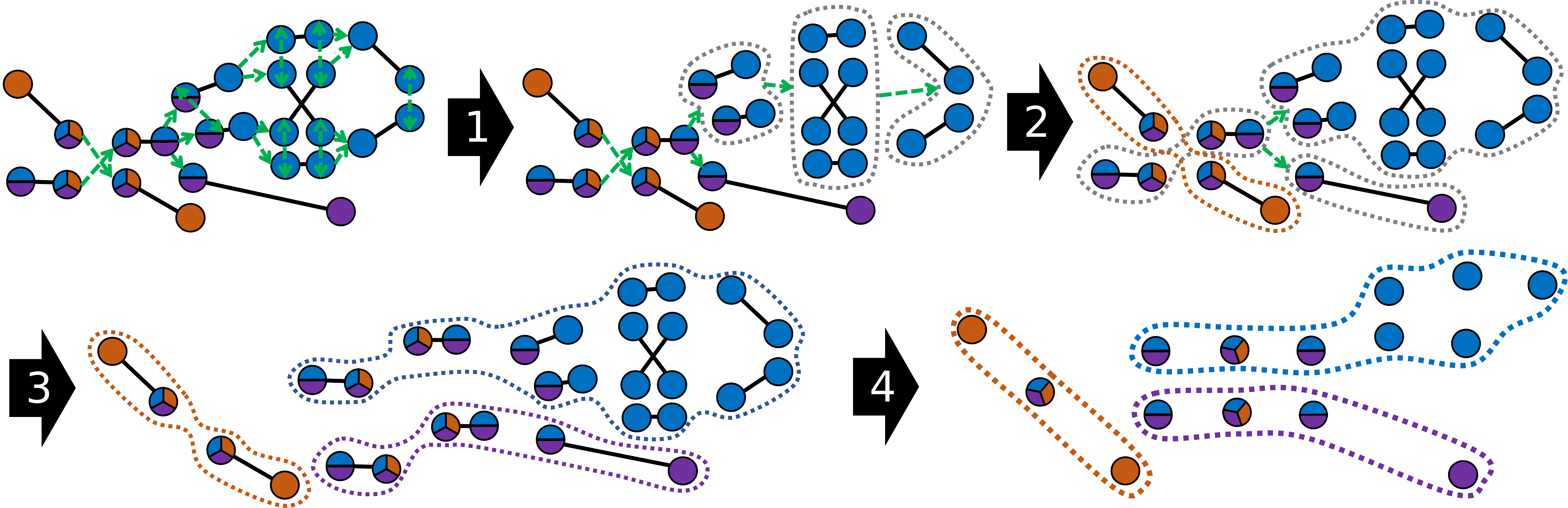}
  \caption{Illustration of the four phases of track construction from the purified edge graph using articulations and elbows defined in Section~\ref{pipeline:triplet_building}. (1) Connecting left and right elbows, (2) applying a WCC while omitting articulations with shared edges, (3) designating each residual link as a unique track, and (4) substituting edges with hits.}
  \label{fig:track_building}
\end{figure}

To build tracks from the purified edge graph $G^{\text{edge}}_{\text{purified}}$, simply applying the WCC algorithm would 
identify sets of connected edges, allowing for the reconstruction of tracks that share exactly one hit. However, electron-positron 
pairs are created with a very small opening angle between the two tracks and therefore frequently share multiple hits at the beginning
of the tracks. In order to account for this, the process of building tracks from triplets involves 4 steps as illustrated in
Figure~\ref{fig:track_building}.

First, the left elbows and right elbows are connected, leaving only the articulations to connect. Duplicate edges resulting from elbow connections are removed, so that when two tracks share their initial hits, it is equivalent to two articulations sharing an edge. Second, a WCC algorithm is applied to the edge graph, excluding 
these shared articulations. The shared articulations now act as connections between two sets of connected edges, with one set 
being shared. Third, a new track is formed for each remaining articulation, effectively duplicating the shared set of connected 
edges. Finally, edges are replaced by their corresponding hits, converting the sets of connected edges into sets of connected 
hits, thereby representing the tracks.

\section{Physics Performance}
\label{sec:performance}

The conventions and definitions for track-finding performance in LHCb are outlined in~\cite{Li:2021oga}. A track is matched to a 
particle when at least 70\% of its hits are associated with that particle, forming a set of \textit{matching candidates} (track, particle). 
The metrics used to assess the track-finding performance are presented in Table~\ref{tab:metrics}. The uncertainties associated with 
track-finding efficiency are determined via the Bayesian method with a uniform prior~\cite{Paterno:2004cb} and computed using 
the \texttt{ROOT} software~\cite{Brun:1997pa}.

\begin{table}[!htb]
\begin{center}
{
\renewcommand{\arraystretch}{1.5}
\begin{tabular}{l|ll}
\hline\hline
Metric & Definition & Formula \\
\hline
Efficiency & Proportion of matched particles & $\frac{\text{\# matched particles}}{\text{\# particles}}$ \\
Clone rate & Proportion of redundant candidates & $\frac{\text{\# candidates - \# matched particles}}{\text{\# candidates}}$ \\
fake rate & Proportion of unmatched tracks & $\frac{\text{\# unmatched tracks}}{\text{\# tracks}}$ \\
Hit efficiency & Average proportion of matched hits per particle & $\left\langle\frac{\text{\# matched hits}}{\text{\# hits on particle}}\right\rangle_{\text{candidates}}$ \\
Hit purity & Average proportion of matched hits per track & $\left\langle\frac{\text{\# matched hits}}{\text{\# hits on track}}\right\rangle_{\text{candidates}}$ \\[2mm]
\hline\hline
\end{tabular}
}
\caption{Metrics for track-finding performance. Efficiency, clone rate, and fake rate encompass all events, while hit efficiency and hit purity are averaged across matching candidates.}
\label{tab:metrics}
\end{center}
\end{table}

Within the VELO's track-finding context, Figure~\ref{fig:tracking_system} illustrates two primary particle categories:
\begin{itemize}
    \itemsep0em
    \item[] \textbf{VELO-only particles}: particles whose tracks are reconstructible in the VELO but not in the SciFi.
    \item[] \textbf{Long particles}: particles whose tracks are reconstructible both in the VELO and in the SciFi.
\end{itemize}
Long track trajectories are bent between the UT and the SciFi due to the magnetic field, which enables momentum measurements. Consequently, reconstructing these tracks is crucial for LHCb physics analyses. The aforementioned categories can be further broken down into three sub-categories:
\begin{itemize}
    \itemsep0em
    \item[] \textbf{No electrons}: All particles except for electrons.
    \item[] \textbf{Electrons}: Only electrons, which are more challenging to reconstruct due to a higher chance of scattering or photon radiation (bremsstrahlung).
    \item[] \textbf{From strange}: Non-electron particles originating from a decay chain with an $s$-quark hadron, excluding electrons. These typically represent tracks originating near the end of the VELO detector, making them harder to reconstruct.
\end{itemize}

The physics performance of the ETX4VELO pipeline is evaluated on a sample of 1000 events. This performance is compared to that of the default search by triplet algorithm used for track finding in the VELO within the Allen framework. The performance metrics for ETX4VELO and the search by triplet algorithm are summarised in Table~\ref{tab:long_performance} for long particles, Table~\ref{tab:velo_only_performance} for VELO-only particles, and Table~\ref{tab:fake_rate} for the fake track rate. These tables also present the performance of the ETX4VELO pipeline without the triplet approach, where tracks are obtained by applying a WCC on the purified graph of hits $G^{\text{hit}}_{\text{purified}}$, as currently implemented in C++/CUDA.

\begin{figure}[!htb]
    \centering
    \includegraphics[width=0.9\linewidth]{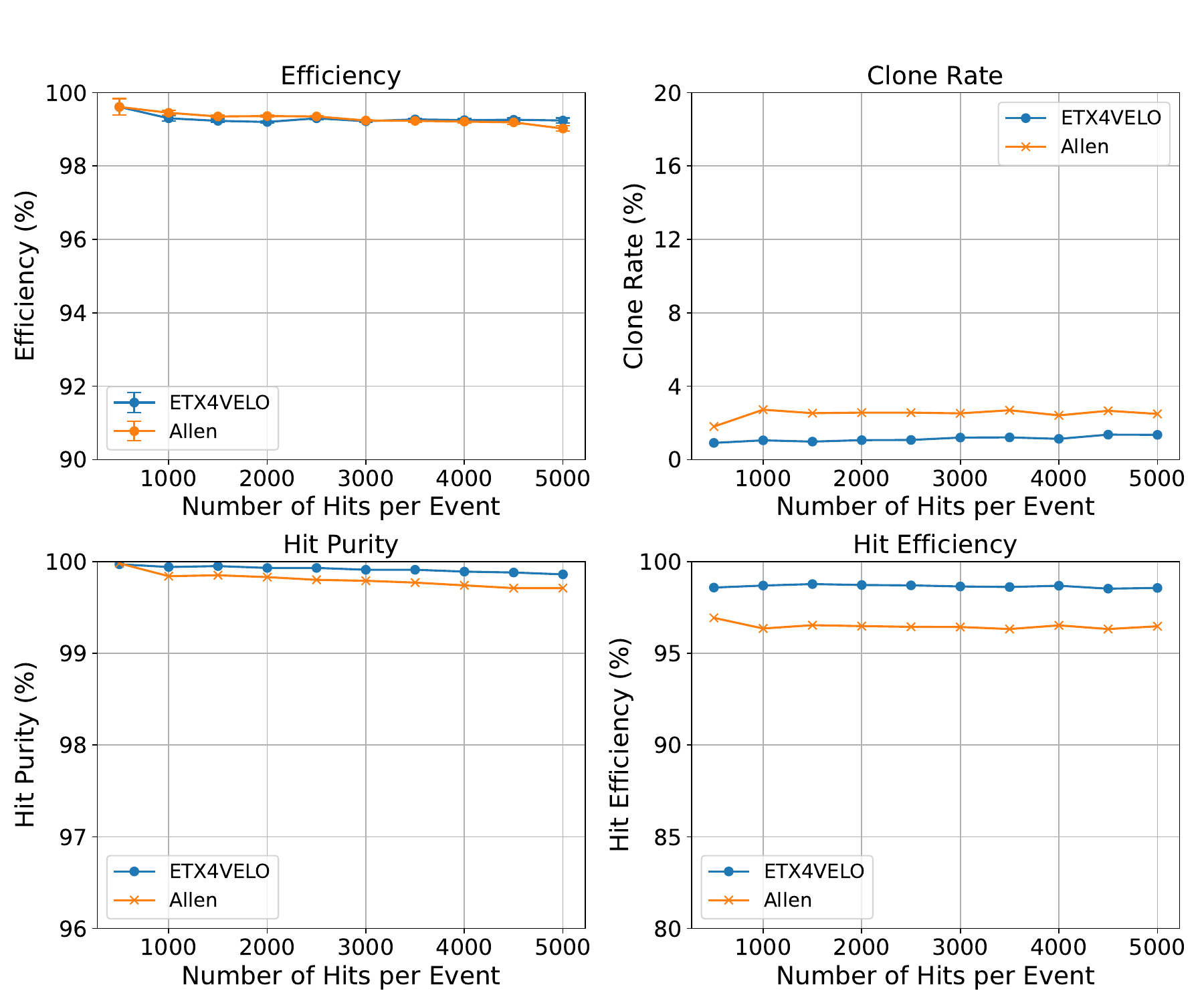}
    \caption{Track-finding performance comparison for the search by triplet algorithm in Allen versus ETX4VELO for long particles, excluding electrons, as a function of the occupancy of the detector.}
    \label{fig:physics-vs-occupancy}
\end{figure}

The physics performance is also compared between the ETX4VELO pipeline and Allen for long particles, excluding electrons, as a function of the occupancy of the detector, i.e. the number of hits in each event, in Fig.~\ref{fig:physics-vs-occupancy}. Various track-finding performance metrics are plotted against the occupancy. Events are split into bins based on their occupancy, and the evaluation of the tracking algorithms is done on the events of each bin. The error bars for the efficiency are binomial errors.  

The ETX4VELO pipeline demonstrates performance comparable to the Allen framework, with a fake rate reduced by more than half. Track quality is significantly better across all categories, with consistently higher hit efficiency and hit purity. It excels particularly in reconstructing electron tracks. However, it performs slightly less well for particles originating from strange decays. This shortfall is likely because these particles tend to be more tilted relative to the beam line and may not have been adequately included during the graph building stage due to the chosen squared maximum distance $d^2_\text{max}$.

\begin{table}[!htb]
\setlength{\tabcolsep}{4pt}
\begin{center}
\begin{tabular}{l|cc|cc|cc|cc}
\hline\hline
Long & \multicolumn{2}{c|}{Efficiency} & \multicolumn{2}{c|}{Clone rate} & \multicolumn{2}{c|}{Hit efficiency} & \multicolumn{2}{c}{Hit purity} \\
\multicolumn{1}{l|}{} & Allen & ETX4VELO & Allen & ETX4VELO & Allen & ETX4VELO & Allen & ETX4VELO \\
\hline
\multicolumn{1}{l|}{No electrons} & 99.35 & 99.35 (97.96) & 2.61 & 1.23 (0.88) & 96.34 & 98.58 (98.42) & 99.78 & 99.92 (99.95) \\
\multicolumn{1}{l|}{Electrons} & 95.21 & 98.10 (51.82) & 3.31 & 3.35 (0.93) & 95.69 & 97.33 (96.46) & 98.37 & 99.55 (95.05) \\
From strange & 97.53 & 97.43 (92.23) & 2.70 & 1.62 (0.61) & 95.85 & 97.95 (96.39) & 99.44 & 99.59 (99.77) \\
\hline\hline
\end{tabular}
\caption{Track-finding performance (in percentages) of the search by triplet algorithm in Allen versus ETX4VELO for long particles. The values in parentheses correspond to the performance of the ETX4VELO pipeline without the triplet approach, as currently implemented in C++/CUDA.}
\label{tab:long_performance}
\end{center}
\end{table}

\begin{table}[!htb]
\setlength{\tabcolsep}{4pt}
\begin{center}
\begin{tabular}{l|cc|cc|cc|cc}
\hline\hline
VELO-only & \multicolumn{2}{c|}{Efficiency} & \multicolumn{2}{c|}{Clone rate} & \multicolumn{2}{c|}{Hit efficiency} & \multicolumn{2}{c}{Hit purity} \\
\multicolumn{1}{l|}{} & Allen & ETX4VELO & Allen & ETX4VELO & Allen & ETX4VELO & Allen & ETX4VELO \\
\hline
\multicolumn{1}{l|}{No electrons} & 97.03 & 97.05 (96.28) & 3.65 & 1.46 (0.87) & 94.07 & 97.68 (97.93) & 99.51 & 99.81 (99.92) \\
\multicolumn{1}{l|}{Electrons} & 67.84 & 83.60 (49.93) & 9.65 & 6.71 (3.51) & 79.57 & 90.83 (85.25) & 97.62 & 99.17 (98.25) \\
From strange & 94.25 & 93.69 (84.33) & 5.16 & 4.09 (1.35) & 90.33 & 97.95 (90.79) & 99.43 & 99.49 (99.72) \\
\hline\hline
\end{tabular}
\caption{Track-finding performance (in percentages) of the search by triplet algorithm in Allen versus ETX4VELO for VELO-only particles. The values in parentheses correspond to the performance of the ETX4VELO pipeline without the triplet approach, as currently implemented in C++/CUDA.}
\label{tab:velo_only_performance}
\end{center}
\end{table}

\begin{table}[!htb]
\centering
\begin{tabular}{l|c|cc}
\hline\hline
 & \multirow{2}{*}{Allen} & \multicolumn{2}{c}{ETX4VELO} \\[1mm]
 &  & With triplets & Without triplets \\[1mm]
 \hline
fake rate & 2.18\% & 1.01\% & 2.07\%\\
\hline\hline
\end{tabular}
\caption{Fake rate of the search by triplet algorithm in Allen versus ETX4VELO, for the full pipeline with triplets and the pipeline excluding the triplet approach, as implemented in C++/CUDA.}
\label{tab:fake_rate}
\end{table}

\section{C++/CUDA Implementation and Throughput Measurement}
\label{sec:gpu-implemenation}

The ETX4VELO pipeline is implemented in C++/CUDA using the Allen framework, the implementation of the first-level of the LHCb trigger on GPUs, in order to measure its computational performance. Our implementation leverages Allen's existing implementations for tasks such as memory management, event loading and dispatching, 
and VELO hit decoding. The classical Allen reconstruction pipeline is benchmarked dispatching 500 events to each of 16 CUDA streams 
and allocating 500 MB of GPU memory per stream. The ETX4VELO pipeline must adhere to these constraints to ensure optimal performance.
Computational throughput is measured using the Nvidia GeForce RTX 2080Ti and GeForce RTX 3090 cards, with 50 repetitions of the pipeline in order to make the
impact of I/O overhead on the throughput measurement negligible.
The following steps of the pipeline have been implemented: (1) embedding network inference, (2) k-NN algorithm, (3) GNN 
inference up to the edge classifier, and (4) WCC algorithm. The implementation of track building from edge triplets is left
for the future, but is not expected to have a major impact on the computational performance of the pipeline.

\subsection{Network Inference}

To infer the embedding network and the GNN trained in PyTorch, inference engines are utilised. We have tried both ONNX Runtime (ORT)~\cite{ort} 
with a CUDA backend and TensorRT (TRT)~\cite{trt}. Both engines require the corresponding PyTorch networks to be exported to the ONNX open-source 
format. For TensorRT, memory allocation can be easily delegated to the Allen framework, making it well-suited for production and 
real-time inference scenarios. In contrast, ONNX Runtime does not readily support this feature. However, ONNX Runtime has the advantage 
of a CPU backend, making it easy to switch the pipeline from GPU to CPU when needed.

For embedding inference, since Allen dispatches 500 events to each CUDA stream, all the hits from these 500 events are provided 
directly to ONNX Runtime and TensorRT, maximizing parallelization within the available memory constraints. For GNN inference, 
events are grouped into batches with a maximum of $2^{20}$ hits and $2^{22}$ edges, which is 
the maximum that GPU memory allows, to accommodate varying event sizes.

Support for the \texttt{scatter\_add} operation during message building posed challenges for both ONNX Runtime and TensorRT. Fortunately, the latest version of ONNX Runtime (v18) supports this operation. However, for TensorRT, a custom plugin had to be implemented. Although TensorRT 10.0 and later versions support \texttt{scatter\_add}, the support is limited to specific type combinations. For example, the operation on INT8 types is not supported.

\subsection{k-NN Implementation}
For each node in plane $p$, a sequential loop iterates over the hits in planes $p + 1$ and $p + 2$ to compute the squared distance in the embedding space. These iterations are performed in parallel for different hits on a single plane and in parallel for different planes. If the distance is below the squared maximum distance $d^2_\text{max}$, the hit is added to the list of $k_\text{max}$ nearest neighbours. If this list is already full, the maximum distance in the list is found and replaced if the new hit is a closer neighbour, though this situation is infrequent. In fact, less than 0.1\% of hits have more that 50 neighbours.

\subsection{WCC Implementation}

A custom implementation of the Weakly Connected Component (WCC) algorithm is used, leveraging the plane structure of the VELO detector.
The purpose of the WCC is to assign a connected component label to each node, indicating the track to which each node belongs.
Each node is initially assigned a unique label, typically its own node index. Next the label of each node in plane $p$ is updated 
in parallel to the lowest label of its connected nodes on the left, sequentially from plane $1$ to plane $n_{\text{planes}} - 1 = 25$ (the planes being numbered from 0 to $n_\text{planes} - 1$). 
If a node on the right is connected to more than two nodes on the left, it will only be updated to connect to one of these two nodes, 
leaving one of the nodes on the left without a proper label. Therefore, it is necessary to repeat the process in reverse, from plane $n_{\text{planes}} - 2 = 24$ to plane 0, considering the connected nodes on the right.

\subsection{Quantization}

For the quantization to INT8 precision of the embedding MLP model, Nvidia's pytorch-quantization\footnote{\url{https://docs.nvidia.com/deeplearning/tensorrt/pytorch-quantization-toolkit/docs/index.html}} library, targeting the TensorRT backend, was used. The 8-bit tensor cores, instead of the standard CUDA cores of an Nvidia GPU, are utilized for matrix-multiplication operations yielding more compute throughput. The quantization parameters of the model were calibrated using 5,000 events. The physics performance of the pipeline using the INT8 version of the embedding is shown in Table~\ref{tab:quantization_performance}. The fake rate is at 1.72\%. For the INT8 case, the rest of the pipeline remains in FP32 precision. The embedding ends up creating roughly an extra 5-10\% of edges and thus dilutes the performance of the pipeline. 

The quantization of the GNN has not yet been achieved due to the \texttt{scatter\_add} operation not being natively supported by TensorRT. The custom plugin for this operation was only implemented for single precision. However, in order to use INT8 quantization for the GNN, the plugin has to support this precision also. The quantization of the GNN currently holds the highest promise for throughput gains.

\begin{table}[!htb]
\setlength{\tabcolsep}{4pt}
\begin{center}
\begin{tabular}{l|cc|cc|cc|cc}
\hline\hline
Long & \multicolumn{2}{c|}{Efficiency} & \multicolumn{2}{c|}{Clone rate} & \multicolumn{2}{c|}{Hit efficiency} & \multicolumn{2}{c}{Hit purity} \\
\multicolumn{1}{l|}{} & INT8 & FP32 & INT8 & FP32 & INT8 & FP32 & INT8 & FP32 \\
\hline
\multicolumn{1}{l|}{No electrons} & 97.66 & 97.96 & 0.77 & 0.88 & 99.95 & 98.42 & 98.23 & 99.95 \\
\multicolumn{1}{l|}{Electrons} & 58.50 & 51.82 & 2.41 & 0.93 & 96.39 & 96.46 & 92.39 & 95.05 \\
From strange & 89.03 & 92.23 & 1.27 & 0.61 & 99.73 & 96.39 & 94.07 & 99.77 \\
\hline\hline
\end{tabular}
\caption{Track-finding performance (in percentages) of the ETX4VELO pipeline for long particles using the FP32 embedding MLP vs the INT8 version. For the INT8 case, the rest of the pipeline remains in FP32 precision.}
\label{tab:quantization_performance}
\end{center}
\end{table}

\subsection{Throughput Results}
\begin{figure}[!htb]
    \centering
    \includegraphics[width=.9\textwidth]{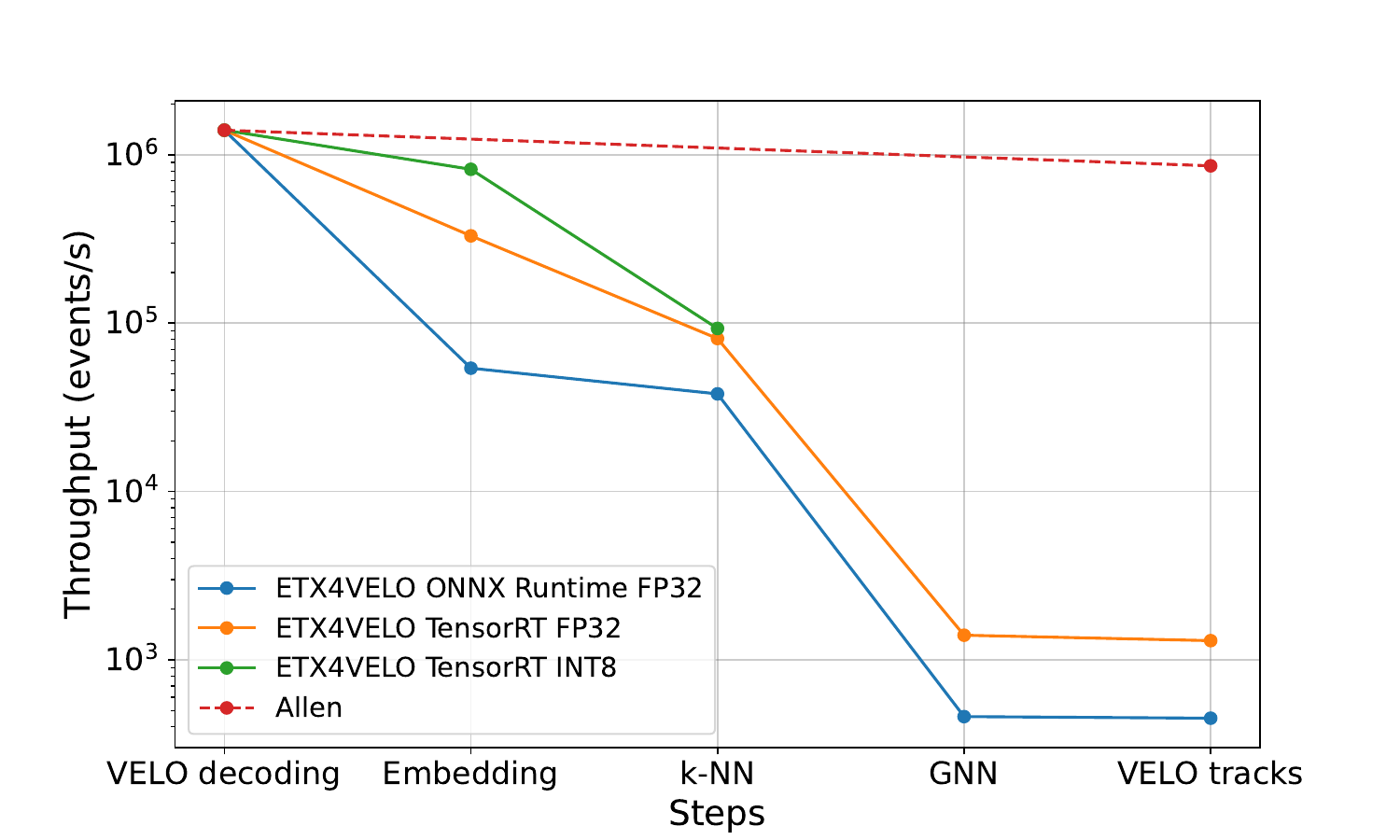}
    \caption{Throughput comparison of track reconstruction in the VELO on an Nvidia GeForce RTX 3090.}
    \label{fig:throughput-comparison-3090}
\end{figure}

The current throughput of the pipeline is shown in Table~\ref{tab:gpu-results-1} and Table~\ref{tab:gpu-results-2} for two different GPU cards. For a visual comparison of the throughput evolution with the algorithm steps on one of these cards, and its comparison to Allen, see Fig.~\ref{fig:throughput-comparison-3090}. The quoted results are from Allen's built-in throughput timer. The throughput is measured after various steps of the pipeline, following the pipeline up to a specific step, hence the ``Up to step'' column. We also give information about the number of streams and the memory used by each stream. The throughput of the ETX4VELO can be separated in 3 columns, ``ORT FP32'', ``TRT FP32'' and ``TRT INT8'', where for each we specify the inference engine and precision that was used for the inference of the ML models.

\par Firstly, the TensorRT implementation in FP32 is significantly faster than the corresponding one in ONNX Runtime. The difference is especially evident in the embedding step, where TensorRT has a throughput of 260k events per second, while ONNX Runtime can process only 46k events per second. Additionally, TensorRT has a lower memory footprint than ONNX Runtime, enabling the GNN to run on multiple streams simultaneously. However, after the k-NN and the GNN, both implementations drop to below 100k, and 1k respectively. On the other hand, the TensorRT implementation with the quantised embedding MLP with INT8 precision, after the embedding step, attains 540k throughput, which is a promising result. After the k-NN however, we see the throughput dropping significantly, namely down to 67k. At present our k-NN implementation is suboptimal as it lacks parallelisation over the neighbours, so there is reason to hope that this throughput loss can be mitigated in the future. 

\begin{table*}[!htb]
\setlength{\tabcolsep}{3pt}
\begin{center}
\begin{tabular}{l|c|c|ccc}
\hline\hline
\textbf{Up to step} & \textbf{\# streams} & \textbf{Memory per stream (MB)} & \multicolumn{3}{c}{\textbf{Throughput (events/s)}} \\
\multicolumn{1}{l|}{} & & & \textit{ORT FP32} & \textit{TRT FP32} & \textit{TRT INT8} \\
\hline
VELO decoding & 16 & 500 & \multicolumn{3}{c}{770k} \\
Embedding & 16 & 500 & 46k & 260k & 540k \\
k-NN & 16 & 500 & 28k & 53k & 67k \\
GNN & 4 (1) & 2,000 (9,600) & 0.32k & 0.86k & - \\
WCC (VELO tracks) & 4 (1) & 2,000 (9,600) & 0.32k & 0.85k & - \\
\hline\hline
\end{tabular}
\caption{Throughput of the GPU implementation of ETX4VELO on Nvidia GeForce RTX 2080Ti. The number of streams and memory used for the GNN and WCC step by the ORT pipeline is shown in parentheses. These throughputs should be compared to 530k for the full Allen pipeline ending in VELO tracks.}
\label{tab:gpu-results-1}
\end{center}
\end{table*}

\begin{table*}[!htb]
\setlength{\tabcolsep}{3pt}
\begin{center}
\begin{tabular}{l|c|c|ccc}
\hline\hline
\textbf{Up to step} & \textbf{\# streams} & \textbf{Memory per stream (MB)} & \multicolumn{3}{c}{\textbf{Throughput (events/s)}} \\
\multicolumn{1}{l|}{} & & & \textit{ORT FP32} & \textit{TRT FP32} & \textit{TRT INT8} \\
\hline
VELO decoding & 16 & 500 & \multicolumn{3}{c}{1,400k} \\
Embedding & 16 & 500 & 54k & 330k & 820k \\
k-NN & 16 & 500 & 38k & 81k & 93k \\
GNN & 8 (1) & 2,500 (9,600) & 0.46k & 1.4k & - \\
WCC (VELO tracks) & 8 (1) & 2,500 (9,600) & 0.45k & 1.3k & - \\
\hline\hline
\end{tabular}
\caption{Throughput of the GPU implementation of ETX4VELO on Nvidia GeForce RTX 3090. The number of streams and memory used for the GNN and WCC step by the ORT pipeline is shown in parentheses. These throughputs should be compared to 860k for the full Allen pipeline ending in VELO tracks.}
\label{tab:gpu-results-2}
\end{center}
\end{table*}

\FloatBarrier
\section{Conclusions}

This work introduced ETX4VELO, a GNN-based pipeline for track reconstruction in the LHCb vertex detector. 
ETX4VELO has the ability to reconstruct tracks sharing hits through an novel triplet-based approach.
When juxtaposed with the default traditional algorithm, ETX4VELO not only matched its efficiency but in some instances, surpassed it. 

The focus has now shifted to deployment, and the GNN-based pipeline has been integrated into Allen. The immediate priority is to optimise the inference time, in order to accommodate the high-rate environment that LHCb operates in. This integration is aiming to provide a comprehensive comparison with the default traditional algorithm. Also, it is expected to offer insights that will guide the refinement of various hyperparameters including the MLP and GNN architectures.
Our pipeline and general network architecture should be applicable to any LHCb tracking algorithm, although the graph construction and eventual quantisation will vary as a function of the occupancy and number of readout channels in the different parts of the detector. 

The code for training and testing ETX4VELO is accessible at \url{https://gitlab.cern.ch/gdl4hep/etx4velo}.
Similarly, the code for the GPU implementation can be found at \url{https://gitlab.cern.ch/gdl4hep/etx4velo_cuda}.

\Acknowledgements
This work is part of the SMARTHEP network and it is funded by the European Union’s Horizon 2020 research and innovation programme, call H2020-MSCA-ITN-2020, under Grant Agreement n. 956086. It is also supported by the ANR-BMBF project ANN4EUROPE ANR-21-FAI1-0011, in collaboration with the Frankfurt Institute for Advanced Studies (FIAS).

The authors extend their sincere gratitude to the LIP6 laboratory for granting access to their versatile GPU cluster, where the majority of the trainings were conducted. We are grateful to the ANN4Europe group at FIAS, led by Ivan Kisel, for their insightful discussions. We thank LHCb's Real-Time Analysis project for its support, for many useful discussions, and for reviewing an early draft of this manuscript. We also thank Roel Aaij and the LHCb engineering teams for the help in setting up the environment dependencies for performing the inference of the models in Allen as well as the LHCb computing and simulation teams for producing the simulated LHCb samples used to benchmark the performance of the algorithm presented in this paper. The development and maintenance of LHCb's nightly testing and benchmarking infrastructure which our work relied on is a collaborative effort and we are grateful to all LHCb colleagues who contribute to it.

\bibliographystyle{JHEP}
\bibliography{biblio}

\end{document}